\documentclass[conference]{IEEEtran}

\usepackage{amsmath,amsfonts}
\usepackage{algorithm}
\usepackage{algpseudocode}
\usepackage{array}
\usepackage[caption=false,font=normalsize,labelfont=sf,textfont=sf]{subfig}
\usepackage{textcomp}
\usepackage{stfloats}
\usepackage{url}
\usepackage{verbatim}
\usepackage{graphicx}
\usepackage{cite}

\usepackage[capitalize,noabbrev]{cleveref}
\usepackage{adjustbox}
\usepackage{pifont}
\newcommand{\cmark}{\textcolor{asparagus}{\ding{51}}}
\newcommand{\xmark}{\textcolor{red}{\ding{55}}}
\usepackage{subcaption}
\usepackage{multirow}
\usepackage{array}
\usepackage{listings}
\usepackage{enumitem}
\usepackage{threeparttable}
\usepackage{xcolor}
\usepackage{booktabs}

\usepackage[acronym]{glossaries}
\newacronym{bram}{BRAM}{Block RAM}
\newacronym{cnn}{CNN}{Convolutional Neural Network}
\newacronym{cpu}{CPU}{Central Processing Unit}
\newacronym{dfg}{DFG}{Data Flow Graph}
\newacronym{dma}{DMA}{Direct memory Access}
\newacronym{dse}{DSE}{Design Space Exploration}
\newacronym{dsl}{DSL}{Domain Specific Language}
\newacronym{dsp}{DSP}{Digital Signal Processor}
\newacronym{fifo}{FIFO}{First In First Out}
\newacronym{fpga}{FPGA}{Field-Programmable Gate Array}
\newacronym{gpu}{GPU}{Graphics Processing Unit}
\newacronym{hbm}{HBM}{High Bandwidth Memory}
\newacronym{hdl}{HDL}{Hardware Description Language}
\newacronym{hls}{HLS}{High Level Synthesis}
\newacronym{ilp}{ILP}{Integer Linear Programming}
\newacronym{iot}{IoT}{Internet of Things}
\newacronym{ir}{IR}{Intermediate Representation}
\newacronym{kpn}{KPN}{Kahn Process Network}
\newacronym{llm}{LLM}{Large Language Model}
\newacronym{llvm}{LLVM}{Low-Level Virtual Machine}
\newacronym{lut}{LUT}{Look Up Table}
\newacronym{miqp}{MIQP}{Mixed Integer Quadratic Program}
\newacronym{ml}{ML}{Machine Learning}
\newacronym{mlir}{MLIR}{Multi-Level Intermediate Representation}
\newacronym{moc}{MoC}{Model of Computation}
\newacronym{nn}{NN}{Neural Network}
\newacronym{qat}{QAT}{Quantization Aware Training}
\newacronym{relu}{ReLU}{Rectified Linear Unit}
\newacronym{sa}{SA}{Systolic Array}
\newacronym{soa}{SoA}{state-of-the-art}

\definecolor{asparagus}{rgb}{0.53, 0.66, 0.42}
\definecolor{vermillon}{rgb}{0.80, 0.40, 0.00}
\definecolor{marine}{RGB}{100, 143, 255}

\begin{document}

\sloppy

\title{MING: An Automated CNN-to-Edge MLIR HLS framework}

\author{
\IEEEauthorblockN{Jiahong Bi, Lars Sch\"utze, Jeronimo Castrillon}
\IEEEauthorblockA{
Technische Universit\"at Dresden\\
Dresden, Germany\\
\{jiahong.bi, lars.schuetze, jeronimo.castrillon\}@tu-dresden.de
}
}

\maketitle

\begin{abstract}
    Driven by the increasing demand for low-latency and real-time processing, machine learning applications are steadily migrating toward edge computing platforms, where \glspl{fpga} are widely adopted for their energy efficiency compared to CPUs and GPUs.
    To generate high-performance and low-power FPGA designs, several frameworks built upon \gls{hls} vendor tools have been proposed, among which MLIR-based frameworks are gaining significant traction due to their extensibility and ease of use.
    However, existing state-of-the-art frameworks often overlook the stringent  resource constraints of edge devices.
    To address this limitation, we propose MING, an \gls{mlir}-based framework that abstracts and automates the \gls{hls} design process.
    Within this framework, we adopt a streaming architecture with carefully managed buffers, specifically designed to handle resource constraints while ensuring low-latency.
    In comparison with recent frameworks, our approach achieves on average $15\times$ speedup for standard \gls{cnn} kernels with up to four layers, and up to $200\times$ for single-layer kernels.
    For kernels with larger input sizes, MING is capable of generating efficient designs that respect hardware resource constraints, whereas state-of-the-art frameworks struggle to meet.
\end{abstract}

\begin{IEEEkeywords}
Hardware Architectures, Compilers, High Level Synthesis, Quantized Neural Network, Edge Computing
\end{IEEEkeywords}

\section{Introduction}
\label{sec:intro}
Resource- and compute-intensive applications such as object recognition demand changes in \gls{iot} infrastructures, that historically relied on communication technologies and cloud computing.
To meet the requirements of low-latency and real-time tasks, an increasing number of applications are shifting towards edge computing, where \glspl{fpga} are emerging as a highly promising hardware platform due to their superior energy efficiency~\cite{2022-fpga-survey, 2022-fpga-energy, 2024-fpga-energy}.

However, unlike programming for CPUs or GPUs, \gls{fpga} programming is inherently challenging.
Coding using a \Gls{hdl} requires expert knowledge and is known to be error-prone, due to its low-level concurrent nature.
To address this, \gls{hls} tools have been the matter of intense research and commercial development over the past decades.
These tools, e.g., AMD's Vitis~\cite{2020-vitis} and Intel's Quartus~\cite{2025-quartus}, generate hardware designs for \gls{fpga} deployment from a high-level software program, e.g., in C/C++.
However, effectively using \gls{hls} tools for high-performance designs still  requires low-level knowledge, expert use of pragma directives, and an intuition of how the way the source code is written impacts the final resource demands of the design.

To further ease \gls{fpga} programming, several frameworks operate at higher levels of abstraction~\cite{2017-finn, 2018-finnr, 2019-heterocl, 2021-autosa, 2022-scalehls, 2024-allo, 2024-scalehls-hida, 2024-pom, 2025-streamhls}.
Despite being presented as general-purpose solutions, most of them implicitly target \gls{ml} workloads, with a particular focus on \glspl{cnn} in the case of FINN~\cite{2017-finn, 2018-finnr}, and \glspl{sa} in the case of AutoSA~\cite{2021-autosa}.
This focus is further reflected in their optimization objectives, many of them concentrate on accelerating the computation of large-scale and deeply nested loops, which are prevalent in modern \glspl{nn}.
These frameworks make significant contributions to the \gls{fpga} design landscape, advancing various aspects such as spatial architecture optimization~\cite{2021-autosa}, polyhedral compilation~\cite{2021-autosa, 2024-pom}, programmability~\cite{2019-heterocl, 2024-allo}, and streaming dataflow design~\cite{2025-streamhls}.
As \gls{ml} models continue to grow in scale and complexity, however, the designs generated by these frameworks often struggle to meet the stringent resource constraints of edge devices.
In particular, the storage of intermediate results can lead to inefficient hardware utilization, which quickly becomes a limiting factor when deploying modern \gls{cnn} workloads onto resource-constrained \glspl{fpga}.

\begin{figure}[h]
    \centering
    \includegraphics[width=\linewidth]{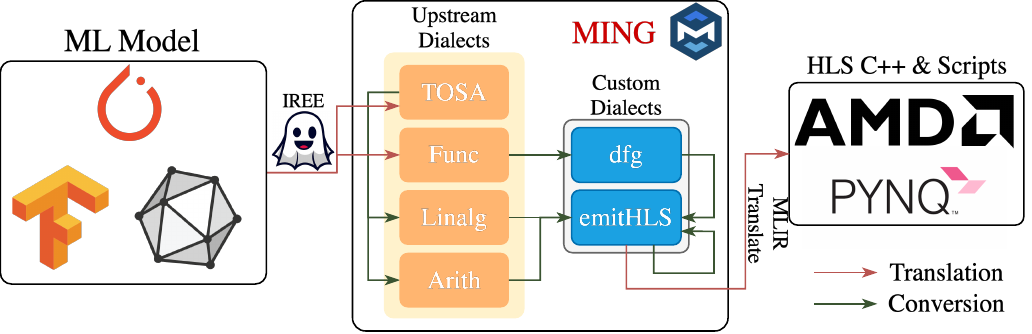}
    \caption{MING Framework Overview}
    \label{fig:ming-overview}
\end{figure}

To address these challenges, we introduce \textbf{MING}, an \textbf{M}LIR-based \textbf{IN}ference \gls{fpga} Hardware \textbf{G}enerator, which performs hardware-aware optimizations tailored for highly resource-constrained edge hardware.
By combining a streaming architecture and resources optimization, MING automatically generates \gls{hls} implementations of the given \gls{cnn} models.
\Cref{fig:ming-overview} depicts an overview of \textbf{MING}, which is built upon the \gls{mlir} infrastructure.
The concrete contributions of MING framework are:

\begin{itemize}[topsep=5pt, partopsep=5pt]
    \item We propose an automated framework that constructs a pure streaming dataflow architecture targeting \gls{hls}-based design, fundamentally differing from state-of-the-art frameworks that rely on \gls{bram}-based interfaces or delegate memory management to the \gls{hls} tool.
    \item We integrate hardware-aware optimizations that respect the resource constraints of edge-level \glspl{fpga} while achieving lower latency.
    \item We propose
    a model to estimate resource utilization  that supports integer arithmetic and is more accurate than models in the state-of-the-art.
    \item We design a lightweight \gls{dse} module within the \gls{mlir} infrastructure to determine the optimal degree of parallelism under resource constraints for the proposed streaming architecture.
\end{itemize}

\section{Related Work}
\label{sec:related}

\begin{table*}[htbp]
    \renewcommand{\arraystretch}{1.2}
    \centering
    \begin{adjustbox}{max width=\textwidth}
    \begin{tabular}{@{}cccccc@{}}
        \toprule
        \textbf{Frameworks} & \textbf{Front-end} & \textbf{Streaming Dataflow} & \textbf{Automatic Pragma Insertion} & \textbf{Adaptive Parallelism} & \textbf{Hardware-aware Optimization} \\
        \midrule
        HeteroCL & DSL & \xmark & \textcolor{orange}{Manual} & \textcolor{orange}{Manual} & \xmark \\
        ScaleHLS & C/C++/PyTorch & \xmark & \cmark & \cmark & \xmark \\
        Allo & Python/PyTorch & \textcolor{orange}{Manual} & \textcolor{orange}{Manual} & \textcolor{orange}{Manual} & \xmark \\
        POM & DSL & \xmark & \cmark & \cmark & \xmark \\
        StreamHLS & C/C++/PyTorch & \cmark & \cmark & \cmark & \textcolor{orange}{Partially} \\
        MING (This work) & ONNX/TensorFlow/PyTorch & \cmark & \cmark & \cmark & \cmark \\
        \bottomrule
    \end{tabular}
    \end{adjustbox}
    \caption{MING comparison with prior frameworks}
    \label{tab:frameworks}
\end{table*}

To ease the translation of \gls{ml} models into \gls{fpga} implementations, numerous compiler frameworks have been proposed by both academia and industry to streamline and accelerate the hardware development process.
We focus in this paper on the frameworks that are capable of transforming \gls{cnn} models into hardware designs.

FINN~\cite{2017-finn,2018-finnr}, initially proposed by AMD, adopts a hybrid template-based approach that combines \gls{hdl} and \gls{hls} methodologies to accomplish this task.
On one hand, the \gls{hdl} templates provide highly efficient implementations of instances, such as data buffers and control logic;
on the other hand, the \gls{hls} templates leverage the rich set of utilities, functions, and pragmas offered by \gls{hls} to simplify and accelerate kernel design.
Similar design methodologies can be found in hls4ml~\cite{2021-hls4ml}, SAMO~\cite{2022-samo} (a FINN-based framework), and ONNX2MDC~\cite{2025-onnx2mdc}.
In addition, DOSA~\cite{2023-dosa} offers a more flexible approach by selecting and combining different templates for \gls{nn} operators from various frameworks, such as FINN, hls4ml and SAMO, to achieve more optimized hardware designs.

However, when encountering novel or unconventional computational kernels for which predefined templates do not exist, designers are required to design them manually.
Consequently, a variety of code generation–based frameworks have been developed to provide more design automation.
These frameworks eliminate the need for manually crafting templates for specific kernel types and instead employ compiler techniques to automatically rewrite and optimize the source code.

HeteroCL~\cite{2019-heterocl} and Allo~\cite{2024-allo} both introduce \gls{dsl}-based approaches that offer a user-friendly programming interface while enabling the generation of high-performance code tailored to user-specified optimization directives.
However, these approaches still offload most of the optimization effort to the users, demanding a considerable level of hardware expertise from them.
To overcome this persisting limitation, several automated frameworks have been proposed, such as ScaleHLS~\cite{2022-scalehls, 2024-scalehls-hida}, which applies graph-level pipelining optimizations, and POM~\cite{2024-pom}, which adopts polyhedral optimization techniques similar to those employed in Polly~\cite{2012-polly}.
StreamHLS~\cite{2025-streamhls} adopts a streaming dataflow paradigm to enhance data transfer efficiency between dataflow nodes and to ensure deterministic execution behavior, owing to the intrinsic characteristics of streaming communication.
StreamHLS further introduces an automated \gls{dse} framework that optimizes stream utilization while simultaneously adhering to the hardware constraints imposed by \gls{dsp} resources.

Other frameworks tailored for specific scenarios, such as AutoSA~\cite{2021-autosa}, which generates \gls{sa} architectures on \gls{fpga};
the work in~\cite{2023-mlir-adaptor}, which adapts various versions of \gls{llvm} \gls{ir} for compatibility with the Vitis HLS tool;
Sisyphus~\cite{2025-sisyphus} employs \gls{llm}-based prompt understanding to automatically insert appropriate Merlin pragmas~\cite{2023-merlin} into C/C++ source code;
and StreamTensor~\cite{2025-stream-tensor}, which generates energy-efficient hardware for \glspl{llm} using a custom tensor representation, are beyond the scope of this work and will therefore not be discussed further.

\Cref{tab:frameworks} presents a comparative overview of several aforementioned code generation–based frameworks, all of which are built upon the \gls{mlir} compiler infrastructure~\cite{2021-mlir}.
To some extent, all the frameworks listed in the table are capable of optimizing \gls{hls} code for \gls{fpga} hardware by automatically inserting compiler pragmas and applying adaptive loop-level parallelism.
These approaches achieve low execution time on high-end \gls{fpga} devices, which are increasingly employed in cloud computing services and feature abundant hardware resources.
However, these frameworks overlook the limited on-board hardware resources, especially \gls{bram}, including StreamHLS, leading to suboptimal or even infeasible hardware deployments while deploying to edge devices.

\section{Background and Motivation}
\label{sec:bg}

\subsection{Motivating Example}

\begin{figure}[t]
    \centering
    \subfloat[Dataflow StreamHLS]{%
        \includegraphics[width=\linewidth]{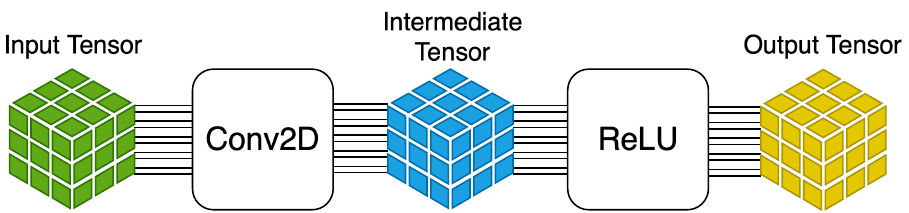}
        \label{fig:motivation-streamhls}
    }\\[1ex]
    \subfloat[Dataflow MING]{%
        \includegraphics[width=0.85\linewidth]{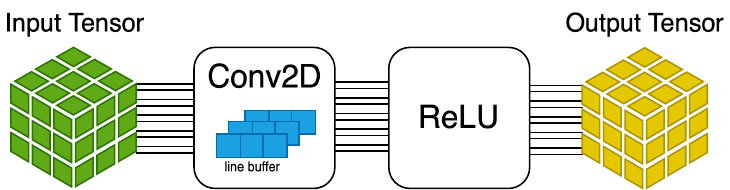}
        \label{fig:motivation-ming}
    }
    \caption{Comparison between StreamHLS and MING}
    \label{fig:motivation}
\end{figure}

\Cref{fig:motivation} gives an intuition of how MING improves the state of the art (StreamHLS in this concrete case) on the basis of a simple \gls{cnn} layer with convolution followed by a ReLU operation. 
As shown in \Cref{fig:motivation-streamhls}, the data from the input tensor (in host memory) are read into multiple streams, which are subsequently used for the computation of the Conv2D operation.
The results are then written into intermediate tensors using a similar streaming mechanism, which are later consumed by the next dataflow node, here, the ReLU operation.
To improve performance, StreamHLS reorders the intermediate tensor into an additional newly created tensor (omitted in the figure for clarity).
This behavior repeats for any subsequent layers cascading after the ReLU operation.

\begin{figure}[ht]
    \centering
    \includegraphics[width=\linewidth]{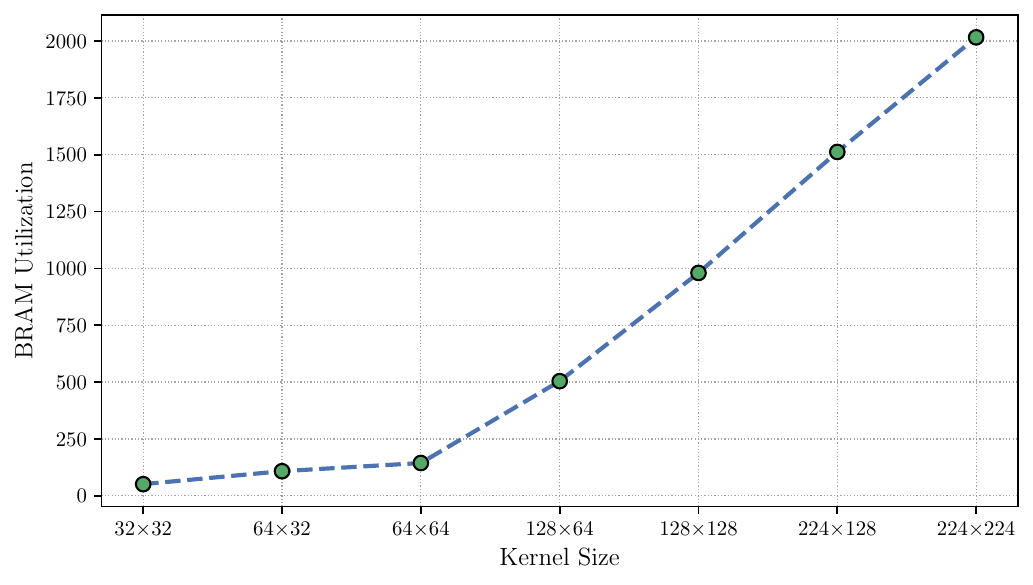}
    \caption{StreamHLS Single-Layer BRAM Utilization}
    \label{fig:streamhls-bram}
\end{figure}

StreamHLS allocates multiple memory instances to store the intermediate results exchanged between dataflow nodes.
In the case of a single Conv+ReLU layer, as the input size scales up, the \gls{bram} utilization increases almost linearly, as illustrated in \Cref{fig:streamhls-bram}.
This trend becomes even more pronounced when scaling up the network depth, where the \gls{bram} utilization continues to grow rapidly with each additional layer.
In contrast, MING eliminates the storage of intermediate data between computation nodes by employing a fully streaming architecture, in which data are produced into a stream and immediately consumed by the subsequent computation node.
The only on-chip memory (\gls{bram}) utilized by MING consists of line buffers, which are instantiated with small sizes in certain node types to temporarily collect data from the input streams.
Further details on the implementation of this mechanism are provided in \Cref{sec:method}.

\begin{figure*}[t!]
    \centering
    \includegraphics[width=\linewidth]{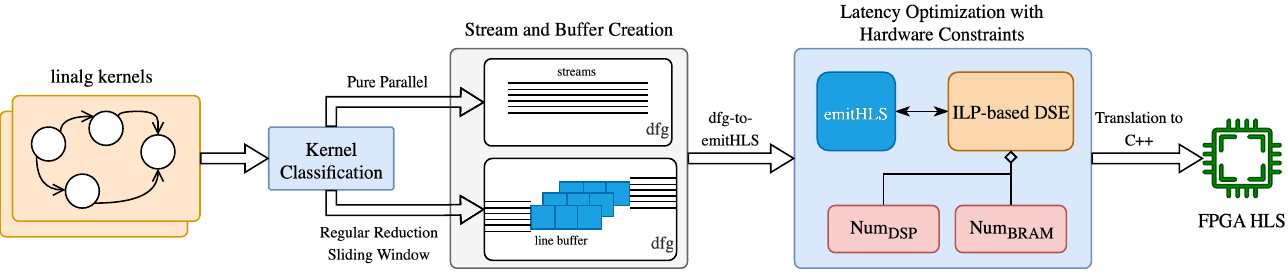}
    \caption{MING Compilation Flow}
    \label{fig:ming-flow}
\end{figure*}

\subsection{MLIR}

\gls{mlir} is a sub-module within the \gls{llvm} framework, designed to provide a comprehensive and reusable compiler infrastructure that supports extensibility and is particularly suited for building \glspl{dsl}.
It has been steadily gaining increasing attention and adoption in both academia and industry.
The core concept of \gls{mlir} is the notion of \textit{dialects}, which represent different abstraction levels and computational domains, for instance, \texttt{TOSA} for tensor operations, \texttt{linalg} for linear algebra computations, and \texttt{arith} for arithmetic operations.
Each dialect comprises a collection of \textit{operations}, \textit{types}, and \textit{attributes} defined for a specific abstraction.

Another fundamental feature of \gls{mlir} is the \textit{pass} mechanism, which can be categorized into two types: \textit{transformations}, applied within the same dialect, and \textit{conversions}, which translate the \gls{ir} from one dialect to another.
A variety of built-in passes are provided to perform traditional compiler optimizations, such as constant folding, common sub-expression elimination, and other canonical transformation techniques.

Motivated by the need for the abstraction of dataflow semantics with strong determinism guarantees for this work, we adopt an existing custom dialect, \texttt{dfg-mlir}~\cite{2024-dfg}, which is grounded in the dataflow \gls{moc}.
This kind of \gls{moc} facilitates an intuitive description of parallelism and supports a wide range of analyses and optimizations~\cite{2023-moc}.
Within the \gls{mlir} infrastructure, the \texttt{dfg} dialect explicitly captures \gls{kpn} dataflow primitives such as processes and \gls{fifo} channels.
In MING, each \texttt{linalg.generic} operation containing a computation kernel is encapsulated as a dataflow node for a given input.
For example, in a single-layer kernel as shown in \Cref{fig:motivation-ming}, the Conv2D and ReLU kernels are represented as two distinct nodes connected through streaming channels.

Vitis \gls{hls} abstracts dataflow communication using the custom class \textit{hls::stream}, which provides the methods \textit{read} and \textit{write} for interacting with streaming data channels.
Within this framework, we introduce a new dialect, \texttt{emitHLS}, which captures both standard C++
and dataflow-oriented syntax in a \gls{hls} design.
This dialect is converted from \texttt{tosa}, \texttt{func}, \texttt{linalg}, \texttt{arith}, and \texttt{dfg} dialects through a series of conversion passes, as illustrated in \Cref{fig:ming-overview}.
Moreover, we leverage the translation mechanisms provided by \gls{mlir}, such as its capability to translate into \gls{llvm} \gls{ir}, to implement our own translation from \texttt{emitHLS} dialect to \gls{hls} code.

\subsection{Hardware-Aware Optimizations}

When writing \gls{fpga} designs in an \gls{hdl}, developers can explicitly control specific hardware resources, for instance, configurable \glspl{lut}, \glspl{bram}, and \gls{dsp} units.
This fine-grained control enables more efficient utilization of on-board resources and requires a high level of expertise.
Similarly, in \gls{hls} C++, the use of compiler directives via pragmas applied to variables, functions, and loops allows designers to guide the HLS tool in allocating particular on-chip resources.
Such resource-aware optimization plays a crucial role in generating high-performance and energy-efficient hardware implementations.

Here we highlight several essential techniques and pragmas that are widely used in \gls{hls} design and are extensively applied in this work to regulate hardware resource utilization:

\begin{itemize}[topsep=5pt, partopsep=5pt]
    \item \gls{fifo} streams are fundamental to enable efficient producer–consumer communication between computation kernels. The \texttt{STREAM} pragma is used to declare variables as streams with appropriate depth.
    \item Loop-level optimizations, including unrolling and pipelining, are critical for achieving high performance in \gls{hls} design, as they significantly increase parallelism and throughput. The pragmas \texttt{UNROLL} and \texttt{PIPELINE} control these optimizations  with specified unroll factors and initiation intervals (II).
    \item Task-level pipelining applies pipelining optimizations at the graph level, allowing each node in a dataflow graph to execute in a pipelined fashion. This is achieved by inserting the \texttt{DATAFLOW} pragma.
    \item Array partitioning plays a crucial role in \gls{hls} designs with loop unrolling optimizations. Insufficient partitioning can lead to memory access conflicts in unrolled loops, resulting in severe performance degradation. To address this, the \texttt{ARRAY\_PARTITION} pragma is used with specified partition dimensions and types.
    \item Selecting appropriate storage types for arrays and binding them to specific memory resources are essential for providing fine-grained control over memory utilization. In MING, this is achieved automatically through the insertion of the \texttt{BIND\_STORAGE} pragma.
\end{itemize}

\section{MING Optimization Flow}
\label{sec:method}

In previous frameworks, even when employing streaming architectures to transfer data between computation nodes, the presence of large intermediate arrays still consumes a substantial amount of hardware resources.
The partitioned arrays are often implemented using \gls{bram} resources by \gls{hls} tools, which can ultimately lead to designs that cannot be synthesized when targeting resource-constrained edge hardware.

To address this limitation, we propose a fully streaming architecture between computation nodes, in which large intermediate arrays are never materialized in the first place.
Instead, only streams and small line buffers are employed for inter-node communication, enabling highly efficient data transfer while maintaining strict data ordering semantics.

\Cref{fig:ming-flow} illustrates the key stages of the MING compilation flow, where the proposed optimizations are applied.
For each dataflow node, MING first classifies the input \texttt{linalg} operation into a specific kernel type.
Based on this classification, appropriate streams and buffers are instantiated through an analysis of the input kernel’s characteristics.
The kernel is then transformed into the proposed streaming architecture in \texttt{dfg} dialect, which is later then converted to \texttt{emitHLS} dialect.
Finally, leveraging the results produced by the \gls{dse}, MING generates the final optimized implementation.
In this section, we provide a detailed walkthrough of these steps.

\subsection{Kernel Analysis}

Frameworks listed in \Cref{tab:frameworks} adopt the \texttt{affine} dialect as their core optimization layer, leveraging its explicit loop structures to enable polyhedral analyses and transformations such as tiling, unrolling, and dependence checking.
Instead, our approach builds upon the \texttt{linalg} dialect.
The \texttt{linalg.generic} operation provides richer semantic information, explicitly distinguishing parallel and reduction dimensions and maintaining the structural relationship between computation and data.
By operating at this abstraction level, we retain semantic expressiveness while still being able to lower into affine-style loop constructs.

For \texttt{linalg.generic} operations with different computational semantics, the resulting dataflow nodes in MING differ not only in their internal computation patterns but also in the created streaming architectures.
In particular, operations that can be executed fully in parallel exhibit distinct data movement characteristics compared to those that require small on-chip memories to buffer input data, such as kernels with sliding-window behavior.
In MING, \texttt{linalg.generic} operations are therefore categorized into three kernel types: pure parallel, regular reduction, and sliding window, each of which is associated with specific dataflow and buffering strategies to best exploit the underlying hardware parallelism.
This section presents the analysis process used to determine the category of each \texttt{generic} kernel, serving as a prerequisite for transforming them into hardware-friendly designs.

\begin{figure}[ht]
\centering
\includegraphics[width=\linewidth]{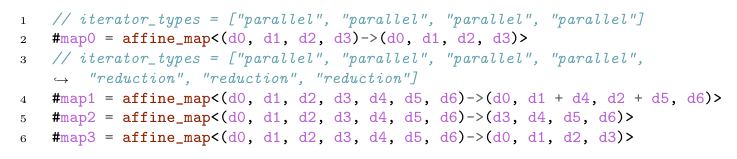}
\caption{Indexing Maps Example}
\label{lst:index-map}
\end{figure}

\Cref{lst:index-map} shows an example of indexing maps used in the \texttt{linalg.generic} operations to be transformed.
An affine map represents all the dimensions, the input affine expressions, involved in the operands and their mappings to the loop indices, which are expressed as affine functions.
The iterator types vector defines the iteration behavior of each dimension.
As their names suggest, a \textit{parallel} iterator operates independently without self-dependence, whereas a \textit{reduction} iterator exhibits reduction behavior, such as that of an accumulator, which updates its own value across iterations.

For example, \textit{map0} in \Cref{lst:index-map} contains four input dimensions, all of which are of the \textit{parallel} type, and each dimension is mapped to itself, forming what is known in \gls{mlir} as an \textit{identity map}.
For \texttt{linalg} operations with only this type of affine map, it is evident that the computations within can be executed fully in parallel across all dimensions (pure parallel).
This high-level semantic enables the transformation into a set of nested loops containing the original computation body (known as payload in \gls{mlir}) after transformation.

\begin{algorithm}[ht]
\caption{Sliding Window Detection}
\label{algo:sliding-window}
\begin{algorithmic}[1]
\Require \texttt{linalg::GenericOp} $op$
\Ensure $(\textit{isSlidingWindow}, \textit{stride}, \textit{dilation})$
\State \textbf{if} all iterators in $op$ are \texttt{parallel} \textbf{then} \Return $(\textbf{false},0,0)$
\For{each input indexing map $M$ of $op$}
  \For{each result expression $E$ in $M$}
    \State Try to rewrite $E$ as $A+B$, where each term is $(\textit{iterator} \cdot \textit{const})$ with const default $1$
    \State Let $A$ attach to iterator $i_a$ with coeff $c_a$, and $B$ to $i_b$ with $c_b$
    \If{one of $i_a,i_b$ is \texttt{parallel} and the other is \texttt{reduction}}
      \State $\textit{stride} \gets$ coefficient of the \texttt{parallel} iterator; \quad $\textit{dilation} \gets$ coefficient of the \texttt{reduction} iterator
      \State \Return $(\textbf{true}, \textit{stride}, \textit{dilation})$
    \EndIf
  \EndFor
\EndFor
\State \Return $(\textbf{false},0,0)$
\end{algorithmic}
\end{algorithm}

The analysis of kernel types becomes more complex when operands are associated with affine maps that include one or more \textit{reduction} dimensions.
As illustrated in \Cref{lst:index-map}, \textit{map1} and \textit{map2} denote the affine maps of the two input tensors, whereas \textit{map3} corresponds to that of the output tensor.
In the case of a \texttt{linalg.generic} operation exhibiting sliding-window behavior, the sliding semantics are implicitly encoded within the affine maps of its input operands.
To distinguish whether a \texttt{linalg.generic} operation with affine maps such as \textit{map1} and \textit{map2} represents a sliding-window or a regular-reduction pattern, we propose a dedicated analysis algorithm, as detailed in \Cref{algo:sliding-window}, which, in the case of identifying a sliding-window operation, additionally extracts key parameters such as stride and dilation.

The key observation is that a sliding-window kernel accesses at least one input tensor using a linear combination of a spatial (\textit{parallel}) iterator and a local accumulation (\textit{reduction}) iterator.
In \gls{mlir}, such access patterns are canonically represented as an affine expression:
\[
E = s \cdot i_p \;+\; \delta \cdot i_r \;
\]
where $i_p$ and $i_r$ denote a parallel and a reduction iterator, respectively.
The coefficient $s \in \mathbb{Z}_{>0}$ is the \emph{stride} (step between adjacent windows along the spatial dimension), and $\delta \in \mathbb{Z}_{>0}$ is the \emph{dilation} (spacing between elements inside a window along the reduction dimension).
Therefore, whenever an input affine map contains a linear combination of exactly one parallel iterator and one reduction iterator with non-zero coefficients $(s,\delta)$, the kernel exhibits sliding-window semantics.
In contrast, regular reduction access patterns will not match this invariant.
The analysis is also lightweight, operating in $O(\sum |E|)$ time, where $|E|$ is the number of map results across the inspected affine maps.

\subsection{Stream and Buffer Creation}

With the classification of \texttt{linalg.generic} operations from the last section, MING then constructs the streaming architecture between computation nodes.
The focus on this step is to reduce array allocations, thereby lowering the overall \gls{bram} utilization.
To enable this key feature, several information must be collected.

To construct the streaming architecture, each \texttt{linalg.generic} operation is analyzed to identify the dimensions within the input and output arrays (or tensors) that can be exploited for parallel processing.
The size of these dimensions determines the number of input and output streams, which are employed to enhance data transfer throughput and overall efficiency.
Subsequently, the content of each computation node is determined jointly by the type of its corresponding \texttt{linalg.generic} operation, the size of the associated streams, and the characteristics of the original computation kernel.

For sliding-window nodes, MING applies the line buffer optimization, a widely adopted technique in \gls{hdl} design.
For instance, in a two-dimensional convolution operation with an input shape of $N \times N$ and a kernel shape of $K \times K$, instead of storing all input data in a large array of size $N^2$, a smaller buffer of size $(K-1) \times N$ is used to retain only a subset of the input lines.
Combined with the newly incoming data line, this buffer provides sufficient data for computing the dot product between the input and the convolution kernel.
Additionally, another buffer with the same shape as the convolution kernel is instantiated to hold the current computation window data.
In our streaming architecture, the line buffer are filled by the data coming from the input streams and then each convolution result is directly pushed into the output streams.
As in other frameworks, particularly those employing polyhedral optimizations, it is difficult to apply this optimization effectively, since loop reordering alters the memory access patterns, thereby disrupting the intended streaming behavior.

For regular-reduction nodes, MING employs a similar strategy by storing the current data line in a small buffer for computation.
The only distinction lies in the absence of the sliding behavior.
When no window movement is involved, the computation is performed directly using this buffer, without the need for an additional window buffer.
Since the implementation closely follows the sliding-window case, further details are omitted here.
For pure-parallel nodes, a \textit{consume–compute–produce} structure is constructed for each data element, where data are consumed from input streams, processed immediately, and the results are produced directly into output streams without intermediate storage.

\begin{algorithm}[htbp]
\caption{Iterator Classification for Stream and Line-Buffer Construction}
\label{algo:stream-buffer}
\begin{algorithmic}[1]
\Require \texttt{linalg::GenericOp} $op$
\Ensure $\mathcal{P}$ (parallel dims), $\mathcal{R}$ (reduction dims), $\mathcal{O}$ (original input dims), $\mathcal{W}$ (window dims)
\State $\mathcal{P},\mathcal{R},\mathcal{O},\mathcal{W}\gets\emptyset$
\For{each input indexing map $M$ of $op$}
  \For{each result expression $E$ in $M$}
    \If{\textsc{is\_single\_dim}$(E)$}
      \If{\textsc{is\_parallel}$(E)$}~~$\mathcal{P}\gets\mathcal{P}\cup E$
      \Else~~$\mathcal{R}\gets\mathcal{R}\cup E$
      \EndIf
    \Else
      \State $\mathcal{O}\gets\mathcal{O}\cup E$
    \EndIf
  \EndFor
\EndFor
\For{each result expression $E$ in output indexing map}
  \If{\textsc{is\_parallel}$(E)$ and $E \notin P$}~~$\mathcal{W}\gets\mathcal{W}\cup E$
  \EndIf
\EndFor
\State \Return $(\mathcal{P},\mathcal{R},\mathcal{O},\mathcal{W})$
\end{algorithmic}
\end{algorithm}

To construct the streaming architecture with optimizations tailored to different computation node types, we introduce \Cref{algo:stream-buffer}.
All relevant information is encoded in the iterator types of dimensions within the affine maps associated with the \texttt{linalg} operation to be transformed, as illustrated in \Cref{lst:index-map}.
The sets returned by this algorithm are structured as follows:

\begin{itemize}
    \item \textbf{Parallel dimensions} ($\mathcal{P}$) represent independent spatial lanes shared by both input and output tensors. Since these dimensions are independent of all others and directly contribute to the output, they are used to define the initial shape of the output streams.
    \item \textbf{Reduction dimensions} ($\mathcal{R}$) correspond to accumulation along specific axes in the input data and, in some cases, also over constant operands involved in the \texttt{linalg.generic} operation. Therefore, these dimensions are used to define the initial shape of the input streams.
    \item \textbf{Original input dimensions} ($\mathcal{O}$) denote the original operand axes that must be preserved for constructing line buffers. These are also the dimensions identified by \Cref{algo:sliding-window} as exhibiting sliding-window behavior.
    \item \textbf{Window dimensions} ($\mathcal{W}$) specify the spatial extent of the sliding window required to construct the compute window, whose data are fetched from the line buffer and subsequently used within the computation kernel.
\end{itemize}

Then, these dimensions and their corresponding sizes are utilized to construct the aforementioned streaming architecture with line buffers and streams for sliding-window and regular-reduction kernels, thereby reducing overall \gls{bram} utilization.
For pure-parallel kernels, which are often positioned after other kernel types, streams of the same size are employed to connect them to their predecessor nodes.

\subsection{Automatic Design Space Exploration}

After analyzing the \texttt{linalg} operations and transforming them into a streaming architecture with controlled hardware utilization using the collected information, MING converts all remaining dialects into the \texttt{emitHLS} dialect, as illustrated in \Cref{fig:ming-overview}, where the arrows directed toward this dialect represent the conversion flow.

Note that the optimizations introduced so far focus solely on reducing hardware utilization.
To improve performance, MING adaptively applies optimizations by automatically insert \gls{hls} pragmas based on the results obtained from an \gls{ilp} model.
The overall design space is significantly reduced due to the strong coupling between loop transformations and hardware resource utilization, resulting in a lightweight \gls{ilp} formulation as shown in \Cref{eq:optimization}.

\begin{equation}
\label{eq:optimization}
\begin{split}
    \min \sum_{v\in\mathcal{V}} Cycles(v_i) &\quad (\mathrm{Objective})\\
    \forall\,\ell \in \mathcal{L}:\quad u_\ell \mid \mathrm{trip}(\ell) &\quad (\mathrm{Unroll\ Constr})\\
    \sum_{d\in\mathcal{D}} \sum_{\ell \in \mathcal{L}(d)} u_\ell\, \eta_{\ell d} \;\le\; D_{\mathrm{total}} &\quad (\mathrm{DSP\ Constr})\\
    \sum_{b\in\mathcal{B}} \sum_{\ell \in \mathcal{L}(b)} u_\ell\, \eta_{\ell b} \;\le\; B_{\mathrm{total}} &\quad (\mathrm{BRAM\ Constr})\\
    \forall\, s \in \mathcal{S}:\ \kappa_{\mathrm{src}(s),s} = \kappa_{\mathrm{dst}(s),s} &\quad (\mathrm{Stream\ Constr})
\end{split}
\end{equation}

The objective of this \gls{ilp} is to minimize the total number of clock cycles required for the entire application.
MING estimates the execution cycles in a manner similar to the Vitis \gls{hls} tools, 
by first counting the number of cycles consumed per loop iteration and then scaling it according to the total iteration count, taking into account the effects of applied loop optimizations.
The key difference from Vitis \gls{hls} lies in the estimation process:
Vitis \gls{hls} derives hardware resource usage from the generated \gls{hdl} code, which is static.
Consequently, any modification or optimization requires restarting the synthesis process to obtain updated estimates.
Compared to StreamHLS, MING explicitly models \gls{bram} utilization and provides a more accurate estimation of \gls{dsp} usage through integer arithmetic, which, as later demonstrated in \Cref{sec:eval}, leads to superior \gls{dsp} efficiency.

At the top-function level, all nodes in the dataflow graph are optimized with task-level pipelining parallelism. 
In a pipeline, the overall latency is ultimately constrained by the finish of the last node in the pipeline.
Therefore, in this model, the total execution cycles are estimated as the sum of the individual node latencies.
The following provides detailed explanations for each of the constraints:

1. \textbf{Unroll Constraint}: Let $\mathcal{L}$ denote the set of loops to be unrolled, and $u_\ell$ represent the unroll factor of a given loop $\ell$.
This is the most straightforward constraint, ensuring that the trip count of each loop is divisible by its corresponding unroll factor.
Another important consideration is that when applying the \textit{PIPELINE} pragma to a loop containing nested loops, the inner loops are automatically fully unrolled.
This behavior arises because the pipelining process must attempt to achieve the specified initiation interval (II), which has implicit influence on the unroll factors of the inner loops.
The result unroll factors are then used in the \textit{UNROLL} pragmas for unrolling the loops.

2. \textbf{DSP Constraint}: Similar to how MING estimates the clock cycles of each node, the \gls{dsp} utilization is modeled by accumulating the \gls{dsp} usage per loop iteration.
Let $D_{\text{total}}$ denote the total number of available \gls{dsp} resources, provided as a compiler argument by the user.
For each loop $\ell$, $u_\ell$ represents its unroll factor, and $\eta_{\ell d}$ denotes the \gls{dsp} consumption per iteration of that loop.
Unrolling a loop by a factor of $u_\ell$ increases its \gls{dsp} usage approximately linearly.
Therefore, the total \gls{dsp} usage can be expressed as the sum of the products $u_\ell \times \eta_{\ell d}$ over all loops, which must not exceed the available resource limit $D_{\text{total}}$.

3. \textbf{BRAM Constraint}: Similar to the \gls{dsp} constraint, the \gls{bram} utilization is also correlated with the unroll factor of the loop in which the variables mapped to \gls{bram} are accessed.
The estimation of \gls{bram} usage is relatively straightforward.
On \gls{fpga} devices, \gls{bram} resources are typically implemented as \textbf{RAM18K} blocks, each capable of storing up to 18{,}432 bits of data.
MING estimates \gls{bram} usage by first calculating the total number of bits required to store the data mapped to \gls{bram}, and then scaling this amount by the corresponding loop unroll factor.
The sum of all such scaled estimates across loops represents the total \gls{bram} consumption, which must be within the limitation set up by the user.
This means that the unroll factor of the loop, in which the variables bound to \gls{bram} are used, are also used in \textit{ARRAY\_PARTITION} pragma for partitioning those variables into data slices to ensure parallel memory accessing.

4. \textbf{Stream Constraint}: Streams are utilized within loops to read data, which is subsequently stored in buffers mapped to \gls{bram} resources.
Moreover, allocating a fixed number of streams without fully exploiting parallelism leads to inefficient utilization of other hardware resources, even though the streams themselves are not implemented in the limited \gls{bram}.
To address this, the number of input and output streams for each node is constrained to match the loop unroll factor of the loop in which the streams are used for reading or writing.
However, since each node is constructed independently and may yield different optimal unroll factors, an additional synchronization constraint is required to ensure consistency, namely, the producer and consumer of each stream must have identical stream sizes.

The solution of this \gls{ilp} yields several key variables, such as the loop unroll factors and the positions where \textit{PIPELINE} pragmas should be applied, which are essential inputs for guiding the subsequent transformation toward the final \gls{hls} design.
Additionally, the estimated clock cycles for the first element to appear in the output stream in each node provide MING with valuable insights for determining appropriate \gls{fifo} buffer sizes.
This estimation helps prevent potential deadlocks, particularly in cases where the dataflow graph contains diamond-shaped structures, such as the residual block as in a \gls{cnn}.

\section{Evaluation}
\label{sec:eval}

In this section, we evaluate the MING framework against three other frameworks: Vanilla as the auto-optimized baseline by Vitis HLS, ScaleHLS~\cite{2022-scalehls, 2024-scalehls-hida} representing designs without streaming architecture and hardware-aware optimizations, and StreamHLS~\cite{2025-streamhls} with its highest optimizations degree as it offers comparable functionality.
All experiments are conducted using the Vitis Suite 2025.1, targeting an edge-level \gls{fpga} platform, the Kria KV260 board~\cite{2025-kria}, on which there are 288 slices of \gls{bram}18K and 1248 \gls{dsp} resources.

\subsection{Experiment Setup}

To evaluate our work we select several commonly used kernels from \gls{cnn} models, specifically those previously evaluated by other frameworks~\cite{2024-scalehls-hida, 2025-streamhls}.
For kernels containing convolution layers, we generate two input sizes --- $32\times32$ and $224\times224$ --- while keeping the remaining dimensions identical.
This setup is designed to emphasize the scalability challenges faced by other frameworks, shown as different utilization of \gls{bram} resources.

In some \gls{cnn} applications, a linear or feed-forward layer is appended at the end of the network, typically characterized by a small number of dimensions but large feature sizes, such as in AlexNet~\cite{2017-alexnet}, for which we additionally design two representative kernels.

At the edge level, \gls{cnn} models used for inference are typically quantized to mitigate the computational overhead associated with floating-point arithmetic, which is particularly expensive on \glspl{fpga}.
For this evaluation, all kernels are quantized to 8-bit integer precision using post-training quantization prior to compilation.
Additionally, to eliminate potential variations caused by different \gls{mlir} versions, we generate identical inputs using IREE, which is capable of producing \gls{mlir} code in the \texttt{linalg} dialect.

These inputs are subsequently compiled into \gls{hls} code, which is then synthesized in the Vitis HLS tool.
After synthesis, we collect key metrics, including the number of clock cycles, \glspl{bram}, and \glspl{dsp} for comprehensive performance and resource utilization analysis.
\gls{dsp} efficiency metric is also used in the evaluation.
It is defined as $E_\text{DSP} = \text{Speedup} / (\text{DSP}_{\text{compare}} / \text{DSP}_{\text{baseline}})$ and reflects the performance gain normalized by the relative \gls{dsp} utilization.

The \gls{bram} and \gls{dsp} usage reported by \gls{hls} closely matches the results obtained after actual hardware synthesis and place-and-route (PnR).
However, other hardware resources, namely LUTs, LUTRAMs, and Flip-Flops (FFs), are often significantly overestimated.
To further investigate the utilization of these resources, we selected kernels of size $32\times32$ that could be successfully synthesized within the resource constraints of the evaluation board.
An additional experiment was then conducted using the \gls{hls} implementation by ScaleHLS, StreamHLS, and MING, covering both hardware synthesis and PnR processes.

\subsection{Results}

\begin{table*}[ht]
    \renewcommand{\arraystretch}{1.2}
    \centering
    \begin{adjustbox}{max width=\textwidth}
    \begin{threeparttable}
    \begin{tabular}{@{}>{\arraybackslash}p{2.5cm}cccc|cccccccccccc@{}}
        \toprule
        \multirow{2}{=}{\textbf{Kernels}} &
        \multirow{2}{*}{\textbf{Input Size}} &
        \multicolumn{3}{c}{\textbf{Vanilla (baseline)}} &
        \multicolumn{4}{c}{\textbf{ScaleHLS@ASPLOS'24}} &
        \multicolumn{4}{c}{\textbf{StreamHLS@FPGA'25}} &
        \multicolumn{4}{c}{\textbf{MING (Ours)}} \\
        \cmidrule(r){3-5} \cmidrule(r){6-9} \cmidrule(r){10-13} \cmidrule(r){14-17}
        & & MCycles & BRAM & DSP & Speedup & BRAM & DSP & $E_\text{DSP}$ & Speedup & BRAM & DSP & $E_\text{DSP}$ & Speedup & BRAM & DSP & $E_\text{DSP}$ \\
        \midrule
        \multirow{2}{=}{Conv+ReLU} & 32x32 & 0.53 & 19 & 5 & 0.74 & \textcolor{teal}{9} & 18 & 0.2 & 1.84 & 51 & 45 & 0.2 & \textbf{504} & 16 & 246 & \textcolor{marine}{10.24} \\
        & 224x224 & 29.2 & \textcolor{vermillon}{707} & 8 & 0.65 & \textcolor{teal}{9} & 35 & 0.15 & 2.06 & \textcolor{vermillon}{2016} & 182 & 0.09 & \textbf{582} & 16 & 246 & \textcolor{marine}{18.92} \\
        \midrule
        \multirow{2}{=}{Cascade Conv Block} & 32x32 & 1.45 & 52 & 10 & 0.74 & \textcolor{teal}{20} & 27 & 0.27 & 2.95 & 116 & 50 & 0.59 & \textbf{44.6} & 32 & 183 & \textcolor{marine}{2.44} \\
        & 224x224 & 86.1 & \textcolor{vermillon}{2280} & 18 & 0.78 & \textcolor{teal}{20} & 52 & 0.27 & 4.06 & \textcolor{vermillon}{6664} & 453 & 0.16 & \textbf{48.6} & 32 & 183 & \textcolor{marine}{4.78} \\
        \midrule
        \multirow{2}{=}{Residual Block} & 32x32 & 1.56 & 89 & 19 & 0.8 & \textcolor{teal}{28} & 53 & 0.29 & 2.02 & 162 & 24 & 1.60 & \textbf{57.8} & 48 & 259 & \textcolor{marine}{4.24} \\
        & 224x224 & 88.6 & \textcolor{vermillon}{3947} & 35 & 0.8 & \textcolor{teal}{28} & 99 & 0.28 & 2.9 & \textcolor{vermillon}{6152} & 127 & 0.80 & \textbf{53.7} & 48 & 259 & \textcolor{marine}{7.26} \\
        \midrule
        Linear & 512x128 & 17 & 265 & 5 & 1.0 & \textcolor{teal}{11} & 10 & 0.5 & \textbf{32319} & \textcolor{vermillon}{6144} & \textcolor{vermillon}{28330} & \textcolor{vermillon}{---} & 125 & 64 & 256 & \textcolor{marine}{2.44} \\
        \midrule
        Feed Forward & 512x128 & 33.9 & \textcolor{vermillon}{463} & 10 & 0.35 & \textcolor{teal}{23} & 16 & 0.22 & \xmark & \xmark & \xmark & \textcolor{vermillon}{---} & \textbf{249} & 96 & 192 & \textcolor{marine}{12.97} \\
        \bottomrule
    \end{tabular}
    \begin{tablenotes}
    \footnotesize
        \item \textbf{Bold text} marks the highest achieved speedup \quad \textcolor{vermillon}{Brown text} marks exceeded resources \quad \textcolor{teal}{Green text} marks the least \gls{bram} usage \quad \textcolor{marine}{Marine text} marks the best DSP efficiency
    \end{tablenotes}
    \end{threeparttable}
    \end{adjustbox}
    \caption{Results derived from HLS tool}
    \label{tab:sim-result}
\end{table*}

\begin{table*}[ht]
    \renewcommand{\arraystretch}{1.2}
    \centering
    \begin{adjustbox}{max width=\textwidth}
    \begin{threeparttable}
    \begin{tabular}{@{}>{\arraybackslash}p{2.5cm}ccccccccc@{}}
        \toprule
        \multirow{2}{=}{\textbf{Kernels} \\($32\times32$)} &
        \multicolumn{3}{c}{\textbf{ScaleHLS}} &
        \multicolumn{3}{c}{\textbf{StreamHLS}} &
        \multicolumn{3}{c}{\textbf{MING (Ours)}} \\
        \cmidrule(r){2-4} \cmidrule(r){5-7} \cmidrule(r){8-10}
        & LUT(\%) & LUTRAM(\%) & FF(\%) & LUT(\%) & LUTRAM(\%) & FF(\%) & LUT(\%) & LUTRAM(\%) & FF(\%) \\
        \midrule
        Conv+ReLU & 11.84 & 4.04 & 8.40 & \textcolor{vermillon}{20.34} & \textcolor{vermillon}{7.02} & \textcolor{vermillon}{14.58} & \textcolor{teal}{9.08} & \textcolor{teal}{1.65} & \textcolor{teal}{5.17} \\
        \midrule
        Cascade Conv & 19.64 & 6.64 & 14.02 & \textcolor{vermillon}{23.46} & \textcolor{vermillon}{7.88} & \textcolor{vermillon}{16.66} & \textcolor{teal}{8.33} & \textcolor{teal}{1.68} & \textcolor{teal}{5.57} \\
        \midrule
        Residual Block & \textcolor{vermillon}{25.38} & \textcolor{vermillon}{8.3} & \textcolor{vermillon}{17.86} & 13.67 & 4.43 & 9.72 & \textcolor{teal}{11.41} & \textcolor{teal}{3.78} & \textcolor{teal}{5.95} \\
        \bottomrule
    \end{tabular}
    \begin{tablenotes}
    \footnotesize
        \item \textcolor{teal}{Green text} marks the least resource utilization \qquad \textcolor{vermillon}{Brown text} marks the most resource utilization
    \end{tablenotes}
    \end{threeparttable}
    \end{adjustbox}
    \caption{Hardware Utilization after Place\&Route}
    \label{tab:pnr-result}
\end{table*}

\Cref{tab:sim-result} presents the results obtained from the \gls{hls} reports generated using the aforementioned frameworks.
As a baseline, the Vanilla implementation exhibits the second-worst performance among all evaluated designs and demonstrates inefficient \gls{bram} utilization for large-size input due to the allocation of memory for intermediate tensors.
When scaling the input size from $32 \times 32$ to $224 \times 224$, the Vanilla implementation consumes over 40$\times$ more \gls{bram} resources.
Moreover, the absence of loop-level optimizations results in minimal \gls{dsp} resource usage as expected.

Surprisingly, ScaleHLS exhibits the worst performance among all evaluated frameworks, approximately 1.5$\times$ slower than the baseline.
An examination of the \gls{hls} code generated by ScaleHLS reveals that, apart from applying pipelining, no additional performance optimizations such as loop unrolling are employed.
This explains the moderate utilization of \gls{dsp} resources in the resulting designs.
Moreover, in several generated dataflow nodes, Write-After-Read (WAR) dependencies are introduced, preventing the \gls{hls} tool from achieving pipelining with an initiation interval (II) of one, thereby limiting the performance.

Compared with the baseline, ScaleHLS does not explicitly allocate new intermediate memories to store data.
Instead, it passes these data directly as function arguments between computation nodes, which are automatically managed by \gls{hls} tool.
In these cases, \gls{hls} implements those arguments as circuit using \gls{lut}, LUTRAM and Flip-Flop (FF).
This explains the minimal \gls{bram} utilization across the results.
However, delegating memory management entirely to the \gls{hls} tool without any specification can also result in infeasible designs.
As shown in \Cref{tab:pnr-result}, the utilization of LUTs, LUTRAMs, and FFs increases most rapidly as the network depth grows.
This eventually leads to a faster exhaustion of the on-board resources.

StreamHLS, in contrast, delivers better performance than both Vanilla and ScaleHLS across all evaluated kernels, achieving approximately a 2$\times$ speedup.
However, when handling larger input sizes, the presence of intermediate data results in massive \gls{bram} allocations which also is prevalent in the baseline.
Furthermore, its automated loop optimization introduces additional memory partitioning, which, as expected, further increases \gls{bram} consumption.
As summarized in \Cref{tab:sim-result}, for larger input sizes, StreamHLS utilizes more than 6,000 \glspl{bram}, which exceeds the limit of \gls{bram} constrain massively.
Even on \glspl{fpga} for the cloud this issue persists when scaling up the input size or increasing the number of network layers.

For kernels involving convolution operations, regardless of the number of layers, StreamHLS successfully generates designs that satisfy \gls{dsp} constraints.
However, for kernels containing linear computations, the framework fails to produce feasible designs, as indicated by the excessive \gls{dsp} utilization shown in the table.
In particular, for feed-forward kernels with multiple cascading Linear layers, the generated design are entirely unsynthesizable in the Vitis HLS tool.

Memory hazards, specifically Write-After-Read (WAR) dependencies, persist in StreamHLS-generated code, causing the same problem observed in ScaleHLS: the \gls{hls} tool cannot achieve an II of one, thus limiting overall performance.
Notably, in the Residual Block kernel, which has more layers than the Cascade Conv Block, the total \gls{bram} utilization is actually lower.
This reduction stems from decreased loop parallelism, leading to fewer memory partitions and consequently more compact memory usage.
A higher parallelism is reflected from higher utilization of LUT, LUTRAM and FF as demonstrated in \Cref{tab:pnr-result}.

MING achieves the overall best performance, delivering an overall speedup of approximately 50$\times$ compared to the baseline, and up to 580$\times$ speedup in the case of single layer kernels.
Owing to the line-buffer and stream-based architecture proposed in \Cref{sec:method}, the generated designs consume substantially fewer \gls{bram} resources than those produced by StreamHLS.
It is worth noting that the utilization of \gls{bram} and \gls{dsp} resources remains consistent regardless of input size, as the same optimization configuration is derived from the \gls{dse} framework.

In certain cases, such as the single-layer Conv+ReLU kernel and the Residual Block, MING utilizes more \glspl{dsp} than other frameworks.
This is attributed to its higher degree of loop-level parallelism, which leads to replicated computational tasks within kernels and consequently greater \gls{dsp} usage, yet with a substantial gain in overall performance while still adhering to hardware constraints.
Additionally, since MING eliminates intermediate arrays with complex indexing for data reads and writes, the proposed streaming architecture enables pipelining with an II of 1, free from any memory hazards.
Additionally, as demonstrated in \Cref{tab:sim-result}, MING achieves the highest \gls{dsp} efficiency among all evaluated kernels.
Moreover, as shown in \Cref{tab:pnr-result}, across all kernels, MING consumes the least LUT, LUTRAM, and FF resources, thanks to the proposed streaming architecture.

The experimental results further validate the fundamental issue described in \Cref{sec:bg}, that is, that existing works do not adequately account for resource constraints.
On cloud-grade high-performance \glspl{fpga}, which feature tens of thousands of \gls{bram} blocks and millions of LUTs, these designs may fit well and achieve good performance through enhanced loop-level parallelization enabled by abundant resources.
However, on more resource-constrained hardware platforms, MING demonstrates significantly better adaptability while maintaining competitive performance without excessive resource overhead.

\begin{table}[h]
    \centering
    \begin{tabular}{cccc}
        \toprule
        \textbf{DSP Constrain} & \textbf{Speedup} & \textbf{DSP} & $\mathbf{E_\textbf{DSP}}$ \\
        \midrule
        1248(100\%) & 504 & 246 & 10.24 \\
        \midrule
        250(20\%) & 19.1 & 76 & 2.25 \\
        \midrule
        50(5\%) & 3.54 & 21 & 0.84 \\
        \bottomrule
    \end{tabular}
    \caption{DSP vs Speedup with Single-Layer $32\times 32$ Kernel}
    \label{tab:dsp-tradeoff}
\end{table}

To further examine the impact of \gls{dsp} resource constraints on speedup and \gls{dsp} efficiency, we configured the \gls{dse} with two smaller \gls{dsp} limits to represent low and extremely low availability of \gls{dsp} resources.
Experiments were then conducted in \gls{hls} using the single-layer kernel with the input size of $32\times32$.
The results are presented in \Cref{tab:dsp-tradeoff}.
Even under extremely constrained \gls{dsp} resources, MING is still able to generate valid designs that stay within the specified limits while maintaining strong speedup compared to the Vanilla baseline.

Although a trade-off exists between the achievable speedup and the number of available \glspl{dsp}, MING consistently outperforms StreamHLS in the same configuration reported in \Cref{tab:sim-result}, achieving higher speedup, lower \gls{dsp} utilization, and superior \gls{dsp} efficiency, even under the most resource-constrained scenarios.
Compared to ScaleHLS, MING achieves both higher speedup and better \gls{dsp} efficiency while utilizing a comparable amount of \gls{dsp} resources in this specific setting.

\section{Conclusion and Future Work}
\label{sec:conclude}

In this paper, we presented \textbf{MING}, an automated \gls{mlir}-based \gls{hls} framework capable of generating \gls{fpga} hardware designs from a wide variety of \gls{cnn} layers.
MING employs a streaming architecture with explicit memory management, substantially reducing hardware resource utilization compared to existing frameworks and thus achieving a much better fit for edge devices without sacrificing performance.
Across all evaluated cases, MING attains up to a 50$\times$ speedup over the most performant state-of-the-art \gls{mlir}-based \gls{hls} framework --- StreamHLS.

As an extension, several optimizations can be incorporated into MING to enhance its generality and performance.
For instance, the code generated by MING already satisfies the design constraints required by the AutoBridge framework~\cite{2021-auto-bridge}, which can further increase the achievable clock frequency, leading to near-linear performance improvement.

For Transformer-like models, the \gls{sa} architecture integrates naturally with MING’s streaming design, opening the possibility of incorporating frameworks such as AutoSA as part of the MING workflow.
This integration would allow AutoSA to benefit from MING’s \gls{dse} mechanism to better meet resource constraints while maintaining high performance.

In the current \gls{dse} module of MING, \gls{fifo} sizes are determined based on the estimated clock cycles of the compute kernels, which generally results in conservative, over-provisioned allocations.
For more accurate and resource-efficient sizing of these streams, tools such as FIFOAdvisor~\cite{2025-fifo-advisor} could be integrated to further reduce hardware utilization.

\section*{Acknowledgment}
This work was supported by MYRTUS. ``MYRTUS is funded by the European Union, by grant No. 101135183. Views and opinions expressed are however those of the author(s) only and do not necessarily reflect those of the European Union. Neither the European Union nor the granting authority can be held responsible for them.''

\bibliographystyle{IEEEtran}
\bibliography{references}

\end{document}